\documentclass[12pt]{article}

\usepackage[utf8]{inputenc}
\usepackage[T1]{fontenc}
\usepackage{lmodern}
\usepackage[margin=1in]{geometry}
\usepackage{amsmath}
\usepackage{amssymb}
\usepackage{booktabs}
\usepackage{longtable}
\usepackage{tabularx}
\usepackage{microtype}
\usepackage{graphicx}
\usepackage{pgfplots}
\pgfplotsset{compat=1.18}
\usepackage[section]{placeins}   

\usepackage[round,authoryear]{natbib}
\usepackage{authblk}

\usepackage{caption}
\usepackage{xcolor}
\usepackage{hyperref}
\hypersetup{
  pdftitle={Does Multi-Agent Debate Improve AI Feedback on Research Papers?},
  pdfauthor={Tomas Havranek and Zuzana Irsova},
  pdfsubject={A pre-registered experiment: authors of 44 economics meta-analyses ranked AI feedback reports on their own papers},
  pdfkeywords={LLM-as-a-judge, meta-science, multi-agent debate, pre-registration, test-time compute},
  colorlinks=true,
  citecolor={blue!55!black},
  linkcolor={blue!55!black},
  urlcolor={blue!55!black}
}

\title{Does Multi-Agent Debate Improve AI Feedback on Research Papers?}

\author[a,b,c]{Tomas Havranek\thanks{Corresponding author:
\href{mailto:tomas.havranek@fsv.cuni.cz}{tomas.havranek@fsv.cuni.cz}. Charles University, Faculty of Social Sciences, 
Institute of Economic Studies, Opletalova 26, 110 00 Prague, Czech Republic.
Project page: \href{https://meta-analysis.cz/debate}{meta-analysis.cz/debate};
pre-registration: \href{https://doi.org/10.17605/OSF.IO/E6XGW}{OSF e6xgw}; supplement:
\href{https://osf.io/7nfyb}{OSF 7nfyb}; replication package:
\href{https://doi.org/10.5281/zenodo.21273528}{Zenodo 10.5281/zenodo.21273528}.}}
\author[a,c]{Zuzana Irsova}

\affil[a]{Charles University, Prague}
\affil[b]{Centre for Economic Policy Research, London}
\affil[c]{Meta-Research Innovation Center at Stanford}

\date{\today}

\begin{document}
\maketitle

\begin{abstract}
\noindent Probably not, at least for meta-analyses in economics. In a pre-registered, identity-masked, within-paper experiment, the authors of 44 meta-analyses ranked three AI reports on their own paper by usefulness for improving it: a single pass by a frontier model against two multi-agent debate tools we built and expected to win. All reports were held to a common length and template. The authors preferred the single pass, by 0.66 rank points over \emph{mad-research} (95\% CI 0.32 to 1.00) and 0.57 over \emph{paper-workshop} (0.16 to 0.95), though \emph{paper-workshop} spent roughly thirty times the tokens. Authors who recalled their journal referee report usually placed it first and never last; in a separate exercise, three AI judges almost always placed the real journal referee report last. Among the three AI reports, Gemini (the judge whose model family wrote none of the reports) would have ranked \emph{paper-workshop} first in the authors' place, reversing the single-pass preference. The reversal warns against substituting an AI judge for the author. We measure perceived usefulness for finished papers; whether AI should referee papers is a separate question.

\end{abstract}



\clearpage

\section{Introduction}\label{sec:intro}

A researcher who wants a second read on a draft could always ask a colleague. Now there are AI
options too, distinguished mainly by how much computation they spend. The cheapest is a
single pass: one capable model reads the draft and returns a report in the time it takes to make
coffee. One step up, two models from different families take turns criticizing the draft and each
other's criticism, and a fresh model writes up the exchange. At the far
end, a workshop of specialized agents argues the paper claim by claim. The bill rises accordingly, from one model call to half a dozen calls to hundreds of thousands of tokens. The computation costs money, but the scarcer input is the author's own time (reading the report and
judging whether the criticism holds up). The question is the
return to that spending, whether it buys feedback an author actually finds more
useful.

This question concerns improving one's own work rather than refereeing someone else's.
\citet{korinek2023} lists feedback on drafts among the ways economists already use language
models, increasingly through agents that chain several model calls \citep{korinek2025agents}, and that is the use we studied: a way for an author to stress-test a paper before
submission, a complement that leaves colleagues and seminar audiences in place.
Whether journals should let models referee is a different question, with incentive problems of
its own \citep{gans2025}; we did not ask it, and no report in this study was used to referee a paper
or fed into any editorial decision.

Whether the extra deliberation helps is contested. Multi-agent debate rests on the claim that models
arguing with each other identify objections a single pass misses, and one strand of evidence supports
it \citep{du2023mad,khan2024debate}; another strand finds that a well-prompted single model does
about as well \citep{smit2023mad,wang2024rethinking}. The disagreement has an economic interpretation. Debate and base-model quality may be substitutes, and so, as
models improve, the cross-check that debate once bought is worth less at the margin
(Section~\ref{sec:conclusion} returns to this).
Both strands mostly score debate on tasks
with checkable answers. But feedback on a research paper has no answer key: the person best
placed to say whether a report helps is the author who would bear the cost of acting on it. What neither
strand has done is put competing AI reports on the same paper in front of that
author and ask which report is most useful. Outside economics, recent work asks whether language
models can generate useful paper feedback or reviews
\citep{liang2024feedback,darcy2024marg,mun2026goodpoint,diwu2026}; in economics,
\citet{pataranutaporn2025} had language models score papers against journal prestige. None of them asks the paper's own author to choose among competing configurations. And \citet{su2025selfrank} show that authors possess private information about the relative
promise of their own papers.

We ran the comparison on 55 economics meta-analyses, 27 our own and 28 by external authors, a
genre chosen for reasons we take up below. We
pre-registered it before generating any report. The three configurations above each wrote a
report on the same paper, the reports were held to a common length and template, and the papers'
authors ranked the three reports by usefulness for improving the paper. The authors saw the
reports in random order, without tool labels. Every paper was already published or accepted, so the rankings measure perceived usefulness in retrospect, not realized improvement. Both multi-agent tools are
ours, and we expected them to win. In our own work we lean on cross-model stress-testing, pushing a
draft past models from different families to catch what a single read misses, and we had found it 
useful enough to build open public tools around the idea. The hypothesis we pre-registered said that at
least one of the two debate tools would prove more useful than the single pass. They did not. The authors ranked
the plain single pass ahead of both multi-agent tools, above \emph{mad-research} on 32 of 44
papers and above \emph{paper-workshop} on 30 of 44 (Section~\ref{sec:results} reports the
margins). A separation this sharp would arise about once in 290 times if the rankings were random within papers, as they would be if the authors had replied just to please us without actually reading the reports (exact permutation $p = 0.0034$). It is also remarkable that the separation is similarly sharp for our own and external papers.

We report three findings. First, more elaborate and more expensive AI feedback brought no detectable gain in the
authors' rankings in this setting. That null is bounded by our design and is not a verdict on multi-agent debate as such. Cost held \emph{paper-workshop} to a deliberately light configuration, blinding held every
report to one length and template, and what the authors ranked is therefore a cheap,
fixed-length version of what these tools produce in normal use. Second, we also had three AI models rank
the same reports (two from the model families that wrote them; one, Gemini, external to all three).
On the real journal referee report the authors and the AI judges disagreed in opposite directions:
authors who recalled where that human report belonged in their ranking usually put it first. When the AI judges ranked it alongside the three tool reports on our own papers, they almost always put it last. Third, the ranking depends on who does it: had the external AI judge ranked the three
reports instead of the authors, it would have placed the most elaborate tool first, reversing the single-pass preference.

The push to register analyses in advance and to make
empirical work in economics transparent and reproducible \citep{christensen2018,ioannidis2025joes}
applies to AI research aids as well \citep{cook2026}. We propose that before such aids become
routine, they be evaluated with pre-specified comparisons and simple baselines, and scored by the intended users, the people who have to act on the output. Parts of the exercise carry beyond this setting. The comparison gives a bounded measure of the return to extra computation in producing fixed-length research feedback. (The machine-learning literature calls that computation test-time compute.) The arms vary the tokens spent on the same task by a factor of
about thirty, and the authors' usefulness rankings did not improve with the spending. And the choice of judge matters: the reversal is a caution for any evaluation that lets a model stand in for the intended user. The protocol that produced both results implements that standard and can transfer to other research aids, including ones whose builders, like us, expect
their tools to win. The authors appear to have judged for themselves rather than delegating to a model, and that is one reason we treat their ranking as the relevant one here. Their orderings agree only weakly with those of all three AI judges, and several told us the reading took real effort (Section~\ref{sec:discussion}).

We chose the meta-analysis genre on purpose, but it is not meant to represent the broader population of economics papers these tools might eventually be used on. The narrowness disciplines the outcome, because a usefulness ranking risks measuring taste unless the reports' claims can be checked. The genre is homogeneous and its methods are relatively standardized. The typical paper collects estimates of one parameter across a set of primary studies, states an inclusion rule, and applies a relatively small number of standard models (such as those available point-and-click at \href{https://easymeta.org}{easymeta.org}). The field also has relative consensus on best practice, including correction for publication bias, codified by the meta-analysis community (the Meta-Analysis of Economics Research Network, MAER-Net) in reporting guidelines and a practitioner's guide \citep{havranek2020guidelines,irsova2024guide,cook2026guidelines}. A report's methodological criticism can therefore be judged against an agreed standard. The \emph{Journal of Economic Surveys} asks the meta-analyses it publishes to follow that guidance, one reason the external papers in our study come from that journal. Suppose a report claims that a meta-analysis fails to correct for publication bias, or that an inclusion rule lets in an incomparable estimate. We can then check the underlying data ourselves and say whether the criticism holds, a check that a one-off empirical paper rarely allows. Using one genre also holds the difficulty of the reviewing task roughly fixed across all 55 papers, so a ranking on one paper is comparable to a ranking on another. Meta-analyses like those in our sample are used to calibrate policy-relevant models and are frequently cited in policy work. They cover quantities such as relative risk aversion, the trade elasticity, the elasticity of intertemporal substitution, the social cost of carbon, the effect of financial incentives on performance, and the effect of class size.

A design in which authors rank reports on their own papers stands or falls with their willingness to read three reports and reply, and unpaid expert time is easier to find close to home. The 27 own papers are the co-authored meta-analyses we have written since 2015 (all available including data and codes at \href{https://meta-analysis.cz}{meta-analysis.cz}), and we serve as associate editors at the \emph{Journal of Economic Surveys}, the journal the 28 external papers come from. The sample itself follows a pre-specified rule. The external stratum is a complete census of every qualifying meta-analysis the journal published from January 2022 onward; a sample dispersed across many journals would have required more discretionary selection. The own stratum adds outlet and vintage variation: several of the own papers appeared in high-ranking journals, among them the \emph{Journal of Labor Economics}, the \emph{Review of Economics and Statistics}, the \emph{Journal of International Economics},  the \emph{Journal of Political Economy Microeconomics}, and the \emph{European Economic Review}. And proximity shows up in neither the response nor the ordering. The cold-emailed external authors answered at an 82\% rate, our own co-authors at 78\%, and the two strata produced the same ordering of the three configurations (stratum-by-arm interaction, permutation $p = 0.97$; Sections~\ref{sec:data} and~\ref{sec:results-rankings} give the accounting).

Section~\ref{sec:tools} describes the design of the comparison and the three configurations.
Section~\ref{sec:data} describes the sample and the recruitment of the authors.
Section~\ref{sec:results} reports the results. Sections~\ref{sec:discussion}
and~\ref{sec:conclusion} state what the results do and do not show.

\section{Study design and feedback configurations}
\label{sec:tools}

We compare three reports per paper in a pre-registered author-ranking exercise. Each paper received
a single-pass report, a \emph{mad-research} report, and a \emph{paper-workshop} Act~I report. The
comparison is between the three configurations as bundles:
the arms share the same base referee prompt (reproduced in the archive) but differ in model family, internal prompting, and number of calls.

The reports take the form of a referee report: a summary, major comments, minor comments, and an
overall assessment. We use the term ``referee report'' for that form only. The authors ranked
feedback on their own papers by how useful it would be for improving them; no report entered a
journal's review process, and none informed an editorial decision. The primary comparison
is three-way, among the AI reports. In a separate four-way comparison on our own papers, the AI
judges ranked one real journal referee report alongside them.

We first describe the common design
and then the three feedback configurations.

\subsection{Design}
\label{sec:tools-design}

To compare what the authors actually saw, we fixed the input, report template, and labels, and
normalized each report to a common length budget, so that format and length did not identify the arm.
A report that argued its case in three times the words would be easy to prefer for reasons that have
nothing to do with the argument. The normalization and blinding pass ran from a fixed specification
applied to all three arms. It rewrote and shortened each report to fit the common template and length budget. When a report complained about garbled equations or OCR (optical character recognition) noise in a paper's extracted text, the pass kept the complaint. We did not manually add or remove substantive
criticism. Because the model re-expressed and shortened the reports, we cannot rule out changes in
emphasis or in which criticisms survived normalization. The
pass made one disclosed exception: a residual
\emph{mad-research} idiom that resisted the blind scrub was rewritten for that arm alone. We rewrote only the phrases that gave the arm away and kept the criticism where we could. Our audit shows the rewrite made the arm harder to detect. This exception cannot account for the single-pass
versus \emph{paper-workshop} contrast, since \emph{paper-workshop} received no such rewrite and was
still ranked below the single pass; it may have affected the \emph{mad-research} contrast, which is
why we disclose it. In the rankings we detected no difference between the two debate tools. The specification and the
script that applied it are part of the replication archive, and a blinding audit is reported in the
supplement.

We removed explicit tool names
and tool-identifying phrasing, presented the three reports in a randomized order, and labeled them
only by position. This is identity masking (residual style fingerprints may have remained).
The papers' authors, who read the reports, knew the reports were AI-generated and produced by
us, but not which report came from which tool.

We fixed this design before generating any report. On 22~June 2026 we registered the study, its
hypotheses, its analysis, and the enumerated 55-paper sample on the Open Science Framework (registration~e6xgw).
The registered hypotheses were H1, that authors rank at least one debate tool's report above the single
pass; H2, that authors prefer one debate tool over the other; H3, that the Gemini judge's ranking of the same three reports concords
with the authors'; and H4, that at least one AI report ranks above the authors' journal referee feedback.
The core design,
primary runs, and pre-specified tests followed that plan, with deviations disclosed below. Registering was an ordinary precaution: with three arms, several outcomes, and our own tools among the things
being judged, a plan written in advance is the only credible guard against reading the results the
way we would have liked them to come out.

\subsection{A single pass}
\label{sec:tools-single}

The first configuration is the obvious baseline: one capable model reads the paper once and writes a
report. We use a single call to a frontier model (Claude Opus 4.8)\footnote{All reports were generated in late June 2026, before the release of GPT-5.6 Sol and before Claude Fable 5 was redeployed; every arm therefore used Claude Opus 4.8 (with GPT-5.5 in the \emph{mad-research} arm).} with a prompt that asks for a
structured report on the paper, the kind of feedback an author could act on before submission. There
is no second opinion and no revision. This is the minimal, and
probably common, way to use a language model for feedback, and it is the cheapest thing one can do:
one model, one call. We refer to it as the single pass.

\subsection{Cross-model adversarial audit}
\label{sec:tools-mad}

The second configuration keeps the single report but subjects it to argument. In \emph{mad-research},
two models from different families, here Claude Opus 4.8 and GPT-5.5 (the latter run through the
Codex command-line tool), run
independent critiques of the paper and then, in an anonymized round, argue with each other's points.
A fresh instance synthesizes the exchange into one report against a fixed severity rubric and keeps a
minority report for objections that did not survive \citep{tool_mad}. Every criticism is tied to a
quotation, so a disagreement has to point at text rather than at a hunch \citep{du2023mad,liang2023mad}.
Using two model families instead of one is deliberate. Errors idiosyncratic to a single model tend to
survive when that model checks its own work, and they plausibly correlate less across families than within them,
so a cross-family audit has a better chance of catching them. The tool automates
a protocol we had run by hand, a two-model duel and a larger multi-agent debate
\citep{tool_duel}. A run makes about six model calls, an order of magnitude more computation than the
single pass.

\subsection{A multi-agent workshop}
\label{sec:tools-workshop}

The third configuration is \emph{paper-workshop}, a Claude-only workshop (all agents Claude Opus 4.8) in which a panel of agents
reviews the paper from several expert viewpoints and argues the contested points, each comment tied to
a quotation and rechecked before it enters the report \citep{tool_workshop}. The tool runs at several
depths; for the comparison here we ran a light desk-review configuration with cross-critique, because
the fuller workshop was infeasible to run 55 times. We evaluate only this report-generating stage
(Act~I). A second act that can return a tracked-changes revision, given the author's source, data, and
code, is a different object from a report and is not evaluated here. Even capped, it is the most
expensive arm, on the order of ten agent calls. Section~\ref{sec:results} puts tokens and dollars on
all three.

A fourth tool of ours, \emph{erc-ai-feedback}, adapts the same report format to grant proposals
\citep{tool_erc}. We mention it for completeness and do not study it here. Table~\ref{tab:tools}
records the mechanisms, call counts, and archival identifiers needed to interpret the comparison.

\begin{table}[htbp]
\centering
\caption{The three feedback configurations add progressively more model calls.}
\label{tab:tools}
\small
\begin{tabularx}{\linewidth}{@{}>{\raggedright\arraybackslash}p{2.5cm}>{\raggedright\arraybackslash}Xc>{\raggedright\arraybackslash}p{3.0cm}@{}}
\toprule
Configuration & Mechanism & Calls & Tool \\
\midrule
Single pass &
One frontier-model call writes a feedback report &
1 &
\emph{(a plain model call, no tool)} \\[0.6em]
mad-research &
Two model families audit the paper and each other; a report is then written from the exchange &
6 &
\emph{mad-research} \\[0.6em]
paper-workshop (Act~I) &
A Claude-only panel reviews the paper from several expert viewpoints and argues the contested points; each comment is tied to a quotation and rechecked &
10 &
\emph{paper-workshop} \\
\bottomrule
\end{tabularx}

\vspace{0.5ex}
{\footnotesize
\begin{minipage}{\linewidth}
\emph{Notes:} Calls are model calls per paper; full token and dollar costs are in
Table~\ref{tab:cost}. The two evaluated tools are archived on Zenodo as \emph{mad-research}
(\href{https://doi.org/10.5281/zenodo.20829175}{10.5281/zenodo.20829175}) and \emph{paper-workshop}
(\href{https://doi.org/10.5281/zenodo.20828996}{10.5281/zenodo.20828996}); the \emph{mad-research}
arm descends from an earlier tool, the research-audit duel protocol
(\href{https://doi.org/10.5281/zenodo.19105954}{10.5281/zenodo.19105954}). A fourth tool,
\emph{erc-ai-feedback} (\href{https://doi.org/10.5281/zenodo.20829165}{10.5281/zenodo.20829165}), adapts the report format to grant
proposals and is not evaluated here. The second act of \emph{paper-workshop}, which returns a
tracked-changes revision rather than a report, is also not evaluated here. Source code for the
tools is on GitHub at \href{https://github.com/tjhavranek}{github.com/tjhavranek}.
\end{minipage}}
\end{table}

\subsection{Deviations from the pre-analysis plan}\label{sec:deviations}

The pre-registered plan (OSF e6xgw) was followed, with these twelve disclosed departures.

\begin{enumerate}
\item The primary hypothesis (H1), that at least one debate arm would be ranked above the single pass, was pre-registered one-sided; the reverse-direction tests reported below were chosen post hoc and are exploratory.
\item The uniform-placement benchmark for the human-referee comparison (H4), under which the human report is equally likely to land in any of the four ranking positions, is our construction, not verbatim in the plan.
\item For own17 and own27 (two of our own papers), our seeded-random choice among the paper's referee reports landed on trivial notes (12 and 60 words), so we substituted the substantive report; the plan added a real referee report but did not specify how one would be selected.
\item Human reports (159 to 2,360 words) were not length-matched to the roughly 1,000-word AI reports; a length-covariate check is in the supplement.
\item Claude and GPT were added as exploratory judges post-registration; the Claude family wrote or co-wrote all three arms and the GPT family co-wrote one, so a self-preference caveat applies, and only Gemini is fully external.
\item The Gemini judge ran on a web ``3.5 Flash'' education account with no-training settings recorded in the run log; we settled on that account after registration and then held it fixed.
\item The \emph{paper-workshop} arm ran a deliberately light configuration of the tool, a desk review with cross-critique. The fuller workshop modes (dozens to hundreds of agents) were infeasible to run 55 times, and Act~II was not evaluated.
\item Paper own16 was accepted on an editor's letter without a proper referee report, so it had no human report to place and the four-way comparison covers 26 papers.
\item Multi-ranker handling (first reply as baseline, keep-all as robustness) was fixed on 25~June 2026, before any reply arrived.
\item We sent no reminders and froze intake at the end of 6~July 2026, matching the invitations' ``within ten days'' wording.
\item The AI judges ranked each paper's reports from the reports plus the paper's opening pages, not its full text; the papers' authors, by contrast, knew their own paper in full.
\item The plan's prose described the exercise loosely as a test of AI ``refereeing''; the paper standardizes on ``feedback,'' since what authors ranked is usefulness for improving their own paper. The outcome and the tests are unchanged.
\end{enumerate}

\section{Data and recruitment}
\label{sec:data}

Our sample consisted of 55 economics meta-analyses: 27 our own and 28 external. The external papers all appeared in the \emph{Journal of Economic Surveys}, which offers a natural, well-defined frame for meta-analyses in economics (we both serve there as associate editors). Appendix Table~\ref{tab:corpus}
lists all 55, with each paper's citation, stratum, and response status.

The external stratum was selected by rule rather than case by case. We screened \emph{Journal of
Economic Surveys} papers that appeared online or in an issue from January 2022 onward and kept papers
that quantitatively synthesized estimates from one empirical literature and had no Havranek or
Irsova author; narrative surveys, methods papers, and guidance papers were excluded. The own
stratum was likewise fixed by rule: the 27 economics meta-analyses we have published or have forthcoming
since 2015 that carry a co-author besides us, all frozen in the registered sample (and all also available including data and codes at \href{https://meta-analysis.cz}{meta-analysis.cz}). The two strata
therefore give us a recent external-journal benchmark plus a set where the probability of getting
author rankings was high. The external benchmark covers the journal's full recent slate of qualifying
meta-analyses, so which papers entered the comparison does not reflect our selection.
Every paper in the sample was already published or accepted when we ran the study, so the three AI reports all bear on finished, accepted work rather than a manuscript still under review.

We picked the meta-analysis genre for the reasons in Section~\ref{sec:intro}: relatively standardized methods
make a report's criticism checkable against the underlying data, and the design could be applied
across all 55 papers (three reports each, one paper-level ranking exercise). The corrections that underpin such checks, such as corrections for publication bias and \textit{p}-hacking, are available in open tooling, including the \href{https://cran.r-project.org/package=maive}{\texttt{maive}} package on CRAN. Scoring the truth of each report's criticisms is not part of the present study.

We put our own 27 papers into the sample to guarantee at
least some responses. Co-authors, we assumed, would answer when asked; a cold-emailed stranger might
not. The guarantee also served speed. We ran and closed the comparison quickly so that the model vintage
would not shift under us while the rankings came in (the frontier model used in every arm already has
a successor, Section~\ref{sec:tools-single}). The estimates are therefore specific to the systems as
they stood in late June 2026.

We emailed every author of each paper for whom we could find an address (each author separately, and excluding ourselves on our own papers) and asked them to read three identity-masked reports on their own paper,
presented in random order as described in Section~\ref{sec:tools-design}, and send a one-line ordering
from most to least useful for improving that paper. Three respondents noted near-ties. We kept their stated order and report a
tie-recode sensitivity (the analysis rerun with those three rankings recoded as ties). The invitation gave a 10-day response window, and we sent no reminder emails.
We froze intake at the end of 6~July 2026, consistent with the ``within ten days'' wording in the
invitations. The invitation also carried one optional request: authors who remembered the journal
referee feedback their paper had received could slot that report into the same ordering, from memory.
Authors of 21 of the 44 covered papers, own and external alike, did so; the recalled-placement
comparison in Section~\ref{sec:results-human} rests on those 21 self-selected recollections.

Table~\ref{tab:sample-accounting} summarizes the accounting. Forty-four of the 55 papers (80\%) drew
at least one ranking. The other 11 drew none; for one of the 11, an author did reply, but only to decline. Across the 44 covered papers we collected 66 valid
rankings from 47 distinct authors. Twenty-six papers gave us exactly one ranking, 15 gave two rankings, two gave three, and one gave four, so 18 papers had two or more independent rankers. Section~\ref{sec:results}
treats the first ranking received for each paper as the baseline and checks robustness to keeping all
66. These 44 first-reply baselines come from 33 distinct authors, since a few responded first on more
than one of their own papers; the one-ranking-per-author robustness check in
Section~\ref{sec:results} uses those 33. We had added our own papers as insurance. Yet the cold-emailed external
authors answered slightly more often than our own co-authors: 82\% against 78\%.

The replies ran from enthusiasm to unease. Some authors found the reports
genuinely useful for revising their own work; others were uncomfortable with the exercise itself. These
are author-side reactions to AI feedback aimed at improving one's own work. They are not evidence for or against
AI refereeing. As Section~\ref{sec:intro} notes, whether journals should let models referee is a separate
question we did not ask. The three example quotations below are anonymous, with identifying details omitted
and a journal name bracketed out. The first declines the exercise, the second weighs the AI reports
against the journal's own referees, and the third bears on what usefulness should mean.

\begin{quote}
``Honestly, both as an author and as a reviewer for various journals \ldots\ I don't feel comfortable
with your initiative.''
\end{quote}

\begin{quote}
``While the AI reports are pretty impressive from a technical perspective, \ldots\ the [journal]
report[s] were more useful to us. The critical points in the AI reports (that are pretty similar)
mostly touch the limitations that we transparently declare in the paper.''
\end{quote}

\begin{quote}
``[F]rom the authors' perspective, the most useful report for improving the paper may also be the
most demanding one.''
\end{quote}

\begin{table}[htbp]
\centering
\caption{The sample produced rankings for 44 of 55 papers.}
\label{tab:sample-accounting}
\small
\begin{tabularx}{0.86\linewidth}{@{}Xr@{}}
\toprule
Quantity & Value \\
\midrule
Papers in the sample & 55 \\
\quad Own meta-analyses & 27 \\
\quad External meta-analyses & 28 \\
Papers with at least one author ranking & 44 \\
\quad Own papers with a ranking & 21 \\
\quad External papers with a ranking & 23 \\
Valid rankings & 66 \\
Distinct authors returning rankings & 47 \\
Papers with two or more rankers & 18 \\
Papers with no ranking & 11 \\
Papers with a recalled placement of the real referee report & 21 \\
Own papers in the AI-judge four-way referee exercise & 26 \\
\bottomrule
\end{tabularx}

\vspace{0.5ex}
{\footnotesize
\begin{minipage}{\linewidth}
\emph{Notes:} The four-way referee exercise
uses 26 own papers because one own paper had no referee report. The baseline
analysis uses the first ranking received for each covered paper; robustness checks keep all 66 rankings.
\end{minipage}}
\end{table}

\section{Results}
\label{sec:results}

We pre-registered the hypothesis (H1) that at least one of the two multi-agent tools would produce a more
useful report than the single pass. Neither did. Across the 44 papers whose authors returned a ranking,
the plain single pass was the one they preferred, on average and in most head-to-head comparisons,
and the two tools we built to improve on it ranked below it.

The four registered hypotheses fared as follows. H1, that at least one debate tool would outrank the single pass, is not supported. The single pass ranked ahead of both tools (Holm-adjusted $p = 0.005$ and $0.026$). H2, that authors would prefer one debate tool over the other, is not supported. The two differ by 0.09 rank points (95\% CI $-0.32$ to $0.50$). H3, that the external Gemini judge's ranking would concord with the authors', is not supported (rank correlation 0.14). H4, that at least one AI report would outrank the authors' journal referee feedback, splits by evaluator. The authors who recalled that feedback usually placed it first and never last, and the AI judges almost always placed it last (Section~\ref{sec:results-human}).

\subsection{Author rankings}
\label{sec:results-rankings}

Table~\ref{tab:rankings} gives each configuration's mean rank and the share of papers on which it was
ranked first (a lower rank means a more useful report). The single pass had a mean rank of 1.59
(95\% CI 1.39 to 1.82) and was ranked first on 55\% of papers. \emph{mad-research} averaged 2.25 and
\emph{paper-workshop} 2.16, and each was ranked first on under a third. The gap is not an artifact
of a few decisive papers: authors placed the single pass above \emph{mad-research} on 32 of 44 papers
and above \emph{paper-workshop} on 30 of 44. Its 55\% first-place share has a wide interval (0.40 to
0.68) that includes one paper in every two. The single pass was not ranked best everywhere, but it was ranked best more often than either tool.

\begin{table}[htbp]
\centering
\caption{Authors ranked the single pass most useful.}
\label{tab:rankings}
\begin{tabular}{lcc}
\toprule
Configuration & Mean rank [95\% CI] & Ranked first (\%) \\
\midrule
\multicolumn{3}{l}{\textit{Pooled (n = 44)}} \\
Single pass    & 1.59 [1.39, 1.82] & 55 \\
mad-research   & 2.25 [2.02, 2.45] & 18 \\
paper-workshop & 2.16 [1.91, 2.41] & 27 \\
\midrule
\multicolumn{3}{l}{\textit{Own papers (n = 21)}} \\
Single pass    & 1.57 & 57 \\
mad-research   & 2.24 & 14 \\
paper-workshop & 2.19 & 29 \\
\midrule
\multicolumn{3}{l}{\textit{External papers (n = 23)}} \\
Single pass    & 1.61 & 52 \\
mad-research   & 2.26 & 22 \\
paper-workshop & 2.13 & 26 \\
\midrule
\multicolumn{3}{l}{\textit{Pairwise contrast (pooled)}} \\
Single $-$ mad      & $-0.66$ [$-1.00$, $-0.32$] & Holm $p = 0.005$ \\
Single $-$ workshop & $-0.57$ [$-0.95$, $-0.16$] & Holm $p = 0.026$ \\
mad $-$ workshop    & $+0.09$ [$-0.32$, $0.50$]  & $p = 0.75$ \\
\bottomrule
\end{tabular}

\vspace{0.5ex}
\footnotesize
\begin{minipage}{\linewidth}
\emph{Notes:} Lower rank is more useful (1 = most useful of three). Baseline
(first-reply) ranking per paper. Single ranked above mad-research in 32 of 44
papers and above paper-workshop in 30 of 44. An exact Friedman test of equal
mean ranks gives $p = 0.0034$. The same ordering appears in both strata. A
permutation test finds no evidence that the arm ordering differs by stratum
($p = 0.97$). Subgroup rows
report point estimates only; interval estimates are reported for the pooled sample, whose precision comes from pooling the strata.
\end{minipage}
\end{table}

Using the table's sign convention, the two pairwise contrasts that carry the result are both negative
and both survive a Holm correction for testing three pairs \citep{holm1979}. Single minus \emph{mad-research} equals $-0.66$
rank points (95\% CI $-1.00$ to $-0.32$, Holm-adjusted $p = 0.005$), and single minus
\emph{paper-workshop} equals $-0.57$ ($-0.95$ to $-0.16$, Holm $p = 0.026$). Negative values mean
the single pass ranked better. The two tools were not distinguishable (H2: mad minus workshop $= 0.09$, 95\% CI $-0.32$ to $0.50$,
$p = 0.75$). The extra calls did not separate them. The interval also bounds the design's resolution: differences between the two tools smaller than about a third to a half of a rank point are not detectable here. A Friedman test \citep{friedman1937} of the three-way
ranking rejects equality (exact permutation $p = 0.0034$). Inference is paper-level throughout:
pairwise contrasts use exact sign-flip permutation tests \citep{ernst2004} on the within-paper rank differences, the
Friedman test permutes ranks within papers, and the 95\% confidence intervals are percentile
bootstraps over papers. Except where a test is flagged as Monte Carlo, these permutation $p$-values are
exact. Four robustness diagnostics use Monte Carlo permutation instead (the stratum-by-arm homogeneity
test, the ranker-clustered sign-flip check, the keep-all Friedman on paper-level mean ranks, and the
tie-recoding sensitivity), and the recalled-placement comparison for the human referee uses exact binomial tests. The pre-registered one-sided hypothesis, that a debate arm would beat the single pass, is not
supported. A one-sided test in the reverse direction is significant, but we chose that direction
after seeing the data and report it only as an exploratory record.\footnote{The exploratory one-sided tests give single ahead of \emph{mad-research}
$p = 0.0008$ and single ahead of \emph{paper-workshop} $p = 0.006$ (exact, one-sided). Because the direction
was chosen post hoc, we do not treat these as confirmatory; the headline is the two-sided effect size.}

We pool the two strata because they tell the same story. Taken alone, our own papers and the
external ones each put the single pass first (own means 1.57 / 2.24 / 2.19; external 1.61 / 2.26 /
2.13), and a test for a stratum-by-arm interaction finds no evidence of heterogeneity (permutation
$p = 0.97$). The ordering is the same in both; the precision of the pooled estimate comes from
combining them. We find it reassuring for the comparison that our own co-authors and cold-emailed strangers ranked the reports the same way. Whatever goodwill the exercise drew, it cannot by itself produce a
particular ordering of reports shown in random order without explicit tool labels.

The \emph{paper-workshop} reports came from a light configuration of the tool (Section~\ref{sec:deviations}),
and all three arms were then standardized to a common length and template. \emph{paper-workshop}'s
roughly 800{,}000 tokens of deliberation per paper (Table~\ref{tab:cost}) were condensed into a report
of about 1{,}070 words, within a couple of dozen words of the single pass's 1{,}096 (word counts are in
the supplement). The rankings order the three configurations under the common report budget; they do not score the
longer, richer output the tools return in
ordinary use.\footnote{These token counts are for the light Act~I configuration we evaluated. Run at
its recommended full settings, with many more expert agents and its second act returning a
tracked-changes revision (Section~\ref{sec:tools-workshop}), \emph{paper-workshop} can plausibly
consume several million tokens per paper, on the order of five to ten million. We did not run that
configuration, and every figure reported here is for the light run.}

\subsection{The cost of a report}
\label{sec:results-cost}

The three configurations differ sharply in what they spend. Table~\ref{tab:cost} sets the counts
side by side. A single pass is one model call and about 27{,}000 tokens; \emph{mad-research} is roughly
six calls and 238{,}000 tokens; \emph{paper-workshop} is on the order of ten agent calls and 800{,}000
tokens. Priced at July 2026 list rates, that is about \$0.20, \$1.87, and \$7.20 per paper, or \$11,
\$103, and \$396 to run all 55. In tokens the most elaborate tool costs about thirty times the
simplest; in API-equivalent dollars, which weight the more expensive model calls, it costs about thirty-five times as much.
(The runs themselves used flat-rate subscription plans, so the marginal cash cost was near zero; the
dollar figures are what the same tokens would cost through the metered API.)

\begin{table}[htbp]
\centering
\caption{The multi-agent reports cost far more to run.}
\label{tab:cost}
\begin{tabular}{lrrrr}
\toprule
Configuration & Calls & Tokens/paper & API-equiv.\ \$/paper & \$ for all 55 \\
\midrule
Single pass    & 1  & 27{,}000  & 0.20 & 11  \\
mad-research   & 6  & 238{,}000 & 1.87 & 103 \\
paper-workshop & 10 & 800{,}000 & 7.20 & 396 \\
\midrule
Ratio (single $=1$) & & 1 : 9 : 30 & 1 : 9.2 : 35 & \\
\bottomrule
\end{tabular}

\vspace{0.5ex}
\footnotesize
\begin{minipage}{\linewidth}
\emph{Notes:} Tokens per paper are rounded run totals. Dollars are
API-equivalent at July 2026 list prices (Claude Opus 4.8 at \$5/\$25 per million
input/output tokens; GPT-5.5, the model the Codex calls ran on, at \$5/\$30); the runs used subscription
plans, so the marginal cash outlay was near zero. Workshop token counts are
run-time telemetry (range 0.76--0.90 million). A normalization pass (about
\$0.19 per paper) was applied equally to all three arms and is excluded from the
tool-only comparison; allocated back equally, it adds about \$0.06 per arm and
leaves the delivered workshop/single ratio at about 28 to 1. Ratios are computed
from unrounded per-token costs.
\end{minipage}
\end{table}

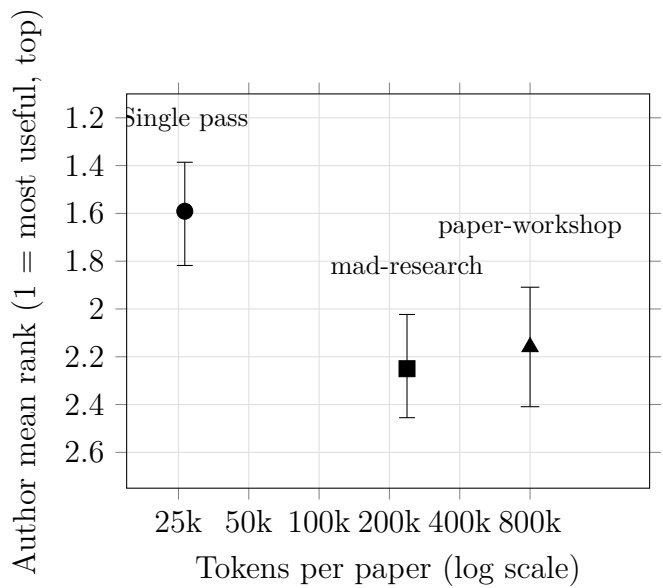
\begin{figure}[htbp]
\centering
\begin{tikzpicture}
\begin{axis}[
  width=8.5cm,
  height=6.8cm,
  xmode=log,
  log basis x=10,
  y dir=reverse,                       
  xlabel={Tokens per paper (log scale)},
  ylabel={Author mean rank (1 = most useful, top)},
  xmin=15000, xmax=2600000,
  ymin=1.10, ymax=2.75,
  xtick={25000,50000,100000,200000,400000,800000},
  xticklabels={25k,50k,100k,200k,400k,800k},
  ytick={1.2,1.4,1.6,1.8,2.0,2.2,2.4,2.6},
  tick align=outside,
  grid=major,
  grid style={gray!25},
]
\addplot[black, only marks, mark=*, mark size=3pt,
  error bars/.cd, y dir=both, y explicit, error bar style={black}]
  coordinates {(26603,1.591) += (0,0.227) -= (0,0.205)};
\addplot[black, only marks, mark=square*, mark size=3pt,
  error bars/.cd, y dir=both, y explicit, error bar style={black}]
  coordinates {(237840,2.250) += (0,0.205) -= (0,0.227)};
\addplot[black, only marks, mark=triangle*, mark size=3.5pt,
  error bars/.cd, y dir=both, y explicit, error bar style={black}]
  coordinates {(800000,2.159) += (0,0.250) -= (0,0.250)};
\node[anchor=south, font=\footnotesize] at (axis cs:26603,1.30)  {Single pass};
\node[anchor=south, font=\footnotesize] at (axis cs:237840,1.90) {mad-research};
\node[anchor=south, font=\footnotesize] at (axis cs:800000,1.75) {paper-workshop};
\end{axis}
\end{tikzpicture}
\caption{Cost and author rankings across the three configurations. Author mean rank (lower is
more useful) against token cost per paper, on a log scale.}
\label{fig:cost}
\end{figure}

Figure~\ref{fig:cost} plots author mean rank against token cost, with cost on a log scale. \emph{mad-research} spent about
\$1.67 more than the single pass and ranked 0.66 places worse, and \emph{paper-workshop} spent about
\$7.00 more and ranked 0.57 places worse. The additional spending did not buy a higher author ranking. Affordability is not the issue here (at these prices all three options are cheap next to an hour of a researcher's time). The question is the return to the extra calls.

\subsection{Robustness}
\label{sec:results-robustness}

The single-pass preference does not depend on how we handle the rankings. Averaging all of each paper's rankings instead of only the first reply, with the resampling
clustered by paper so every paper stays one equal-weight unit, leaves it in place: single
1.60, Monte Carlo Friedman permutation $p = 0.0012$ (the two debate arms trade places under this
weighting, \emph{mad-research} 2.16 and \emph{paper-workshop} 2.24, consistent with their being
indistinguishable). Clustering instead by ranker changes nothing ($p = 0.005$ against \emph{mad-research} and
$0.019$ against \emph{paper-workshop}), and so does keeping only one ranking per author (33 independent rankers, since some authors ranked more than one paper; single is
still ahead of \emph{mad-research}, $p = 0.010$, and of \emph{paper-workshop}, $p = 0.035$). Dropping the one covered paper pre-flagged as atypical (own16,
Section~\ref{sec:deviations}) also leaves the result in place. Either
stratum on its own loses precision but not direction (the stratum-level Friedman tests are
marginal, external $p = 0.068$ and own $p = 0.061$). Pooling is what sharpens them. Recoding as ties the three rankings whose authors said the reports were nearly
indistinguishable does not overturn it either.

Finally, because a fifth of the papers drew no reply, we
computed a worst-case bound on non-response: even if every silent author would have preferred a debate
tool, a majority of the registered 55-paper sample still prefers the single pass to \emph{mad-research} (no-assumption
share bound 0.58 to 0.78) and to \emph{paper-workshop} (0.55 to 0.75). Both lower ends sit above one half. This is the most adversarial assumption about the 11 silent papers, so the ordering does not turn on who chose to reply. The full grids are in the supplement.

Among the 18 papers with more than one
ranker, within-paper agreement was weak (mean pairwise Kendall $\tau = 0.049$), and all rankers
agreed on the first-ranked arm in only three papers. So the paper-level result describes a pattern
across papers rather than agreement among co-authors within one. Individual rankings are evidently noisy. The ordering is
identified by the central tendency across 44 papers, which the ranker-level checks above
show does not hinge on any single ranker. Random disagreement of this kind attenuates the differences between arms rather than creating them, so the ordering that survives the Holm correction does so despite the noise. Several authors volunteered an explanation for the noise:
the three reports tended to raise much the same points, and one wrote that a less strict reader could
reasonably have called them a three-way tie. The extra computation produced reports authors found hard to tell apart, which is itself consistent with the null. The noise does not explain the ordering away, though. The exact permutation test on the full ranking rejects equality at $p = 0.0034$, roughly a 1-in-290 chance of a separation this sharp if the 44 rankings were pure noise within papers. Whether the average ranker read the reports closely or leaned on a model to do it for them is a question the data cannot settle directly. The weak author--Gemini concordance in Section~\ref{sec:results-concordance} ($\tau = 0.14$) argues against wholesale delegation, but we have no per-author measure of how a ranking was actually produced.

\subsection{The AI reports and the human referee}
\label{sec:results-human}

The referee report the paper actually received is a natural, if imperfect, benchmark for an AI report (H4). Here the
authors and the machines disagreed, flatly and in opposite directions. Among the 21 papers whose
authors recalled where their real journal referee report belonged, they ranked that human report first
71\% of the time and never last. We then added the real
referee report as a fourth item alongside the three identity-masked AI reports on our own papers and asked the AI judges to rank
all four. They did the reverse of the authors, who had mostly put the human report first: Gemini ranked it
last on all 26 papers, and Claude and GPT on 25 and 24 of them, giving the human a mean rank near
the bottom of four (Table~\ref{tab:human}).

\begin{table}[htbp]
\centering
\caption{Authors usually ranked the human referee first; the AI judges almost always ranked it last.}
\label{tab:human}
\small
\begin{tabularx}{\linewidth}{@{}>{\raggedright\arraybackslash}Xcc@{}}
\toprule
\multicolumn{3}{@{}l}{\textit{Panel A. Authors' recalled placement (n = 21)}} \\
\midrule
Ranked first                             & \multicolumn{2}{c}{71\% (15/21)} \\
Ranked last                              & \multicolumn{2}{c}{0\% (0/21)}  \\
At least one AI report above the human   & \multicolumn{2}{c}{29\% (6/21)} \\
\midrule
\multicolumn{3}{@{}l}{\textit{Panel B. AI-judge four-way placement, own papers (n = 26)}} \\
\midrule
AI judge & Human ranked last (\%) & Human mean rank (of 4) \\
\cmidrule(lr){1-1}\cmidrule(lr){2-2}\cmidrule(lr){3-3}
Gemini       & 100 (26/26) & 4.00 \\
Claude       & 96 (25/26)  & 3.92 \\
GPT (Codex)  & 92 (24/26)  & 3.85 \\
\bottomrule
\end{tabularx}

\vspace{0.5ex}
\footnotesize
\begin{minipage}{\linewidth}
\emph{Notes:} Authors ranked their real journal referee report first far more
often than any AI report and never last, while the three AI judges almost
always ranked the human report last. The two panels score different objects:
Panel A records the authors' recall of the journal's referee feedback overall,
across referees, on 21 own and external papers; Panel B records the AI judges'
placement of one selected referee report alongside the three identity-masked
AI reports, on 26 own papers. The opposite placements therefore describe
evaluator behavior under two protocols, not opposite verdicts on one report.
The human report always occupied the same final slot (no position rotation),
so the four-way shares are descriptive; human reports were not length-matched
to the AI reports (159--2{,}360 words). Gemini wrote none of the reports, so
self-preference alone cannot explain the pattern.
\end{minipage}
\end{table}

Six qualifications are in order. First, the human report always sat in the same final slot of the
four rather than in a randomized position, so we treat its four-way share as descriptive and do not
lean on a $p$-value. Second, the human reports were not held to the common length budget of the AI
arms (they ran from 159 to 2,360 words). Still, against the authors' recalled first place, the judges placed the human report last in 75 of
the 78 judge-by-paper placements (26 papers, three judges), and the three exceptions concern the
single longest report and one mid-length report, so length alone is unlikely to account for the gap. Together, the first two
qualifications mean the human report was not masked to the standard of the three AI arms: it kept its
own length, format, and a fixed slot. Third, the four-way exercise used real referee reports from our own papers;
we therefore release only their ranks, not their texts, and the benchmark cannot be reproduced from
released referee reports.

Fourth, every paper in the sample is published or accepted, so the recalled referee is
one whose demands the authors ultimately met, and some gratitude may ride along in the memory.
Fifth, the vintages do not match: the journal referee report addressed the manuscript before
revision, while the AI reports and the AI judges worked from the finished paper, so a criticism the
authors have since answered can look obsolete precisely because it was useful.
Finally, the two comparisons score different objects, so the opposite placements are not opposite
verdicts on one report (Panel A is recall across referees; Panel B, one selected report in a fixed
slot among the AI reports). Despite these qualifications, the direction holds. Authors valued the human referee above the
AI reports, and the AI judges valued it below them. Gemini, which wrote none of the reports it ranked,
downranked the human as sharply as the two judges whose families did, so self-preference alone cannot
explain the pattern.

\subsection{Author--machine concordance}
\label{sec:results-concordance}

Setting the human referee aside and returning to the three AI arms, our pre-registered check (H3) asked whether an independent AI judge, here Gemini, would reproduce the authors' ordering. The agreement is weak. The rank correlation between the authors and Gemini is 0.14 (95\% CI $-0.05$ to 0.32, one-sided $p = 0.09$), not distinguishable from zero. The two model families that had
a hand in writing the arms agree with the authors somewhat more (GPT 0.27, one-sided $p = 0.003$; Claude 0.20, one-sided $p = 0.024$). The weak author--Gemini concordance
also cautions against using an AI judge as a substitute for the authors here. The AI judges did share one verdict: all three ranked \emph{mad-research}
last among the three arms, more decisively than the authors, who did not separate it from
\emph{paper-workshop}. The external judge Gemini actually ranked the expensive \emph{paper-workshop}
best (mean rank 1.45, first on 64\% of the 55 papers it judged). Had Gemini ranked the three reports instead of the authors, it would have preferred the most elaborate tool, and the finding would have
reversed. The full judge-by-judge matrices are in the supplement.

\section{Discussion}
\label{sec:discussion}

Interpret the result narrowly. We measured how useful authors found three fixed-length reports
on finished economics meta-analyses. Gains from longer reports, the full workshop, weaker base models, or protocols that vary one design feature at a time remain untested. The genre is a bound of the same
kind. We chose meta-analyses because their claims are checkable, and checkable claims may be where
debate has the least to add; in a messier genre, where readers can reasonably dispute the method, the
same design could return a different answer. So is the sample. The external stratum is a single
journal, the \emph{Journal of Economic Surveys}, where we serve as associate editors, and both strata sit close to us. Authors of meta-analyses in general-interest or field journals, or further from our network, could rank the reports differently, though the own and external strata agree here
(Section~\ref{sec:results-rankings}). The comparison also used one base-model vintage and one single-pass prompt, together with a single implementation of each multi-agent workflow, all run in late June 2026. The result concerns these particular configurations rather than single-pass prompting or multi-agent debate in general. Set against the
multi-agent review literature, the result says where these two debate implementations did not help:
fixed-length author feedback on economics meta-analyses. Where
\citet{darcy2024marg} found a multi-agent pipeline improved generated reviews, we find that authors
did not prefer the multi-agent reports at a common report length, which sits with the strand that has
questioned whether debate reliably beats a well-prompted single model
\citep{smit2023mad,wang2024rethinking,zhang2025overvaluing}.

The result does not grade human referees. The authors usually ranked their journal
referee feedback above the AI reports, and the AI judges, scoring that referee report, almost always ranked it below them. We
report the disagreement and take no side on which ranking is right. Nor is it an argument for AI refereeing. We studied AI feedback as a way to improve one's own work before submission rather than as a substitute for review, and roughly one paper in five drew no ranking, so the observed rankings come from the authors willing to look at AI reports at all. One author declined for exactly that reason: they wrote that, as an author and a reviewer, they did not feel comfortable with the exercise. The discomfort is part of the finding.

We did not observe how the authors formed their rankings, and we did not prohibit AI assistance: an
author could have passed the reports to a chatbot and copied whatever order it returned. Two
observations make wholesale delegation less plausible, though they do not identify the process. First,
the authors' orderings agree only weakly with those of all three AI judges, and with the fully external Gemini least of
all. That is what one would expect if the rankings were their own work rather than a model's verdict relabeled. An author could of course have used a model and still parted ways with our particular external judge, so the weak agreement settles nothing on its own. Second, several authors told us they did
the reading themselves; one wrote that the reports ``often coincided,'' that it ``was not easy to rank
them by usefulness,'' and, plainly, ``I did not use AI,'' then pushed back on our wording: ``Why do you
call it a quick experiment? It took some time to read it.'' Another author, who had expected a quick
task, spent more than two hours on it and was openly skeptical of chatbots; the exercise was unpaid,
and a one-line reply would have satisfied us. The accounts speak for the authors who volunteered them.

The rankings measure perceived usefulness rather than realized improvement. An author judging a report on a
finished paper is telling us which report seems more useful, rather than which one would have made the paper better had it arrived in time. The two can come apart: a report that proposes deeper changes may seem less useful precisely because it asks for more work (one author made the point unprompted;
see the third quotation in Section~\ref{sec:data}). A tone check in the supplement finds
the preferred report only marginally gentler than the others. The single pass was the softest report on only a minority of papers, and
the most action-oriented arm (mad-research) ranked worst, so tone does not by itself account for the rankings.

The
AI-judge rankings carry their own
caveat: two of the three judges belong to model families that helped write the arms (the Claude family
wrote or co-wrote all three, the GPT family one), and LLM judges are
known to favor text from their own family and to reward length and position
\citep{panickssery2024self,saito2023verbosity,wang2023position,zheng2023judge}. Self-preference cannot
be the whole story for the three arms: all three judges, the external Gemini included, ranked
\emph{mad-research} last, and the GPT judge did so even though the GPT family co-wrote that arm. The own-family caveat bears mainly on the human-referee comparison, where the judges downrank
the non-AI text the authors usually put first. Gemini wrote none of the reports and is therefore free of the direct own-family
authorship channel; it agreed with the authors least of the three. That is a caution for what the
evaluation literature calls LLM-as-a-judge, the increasingly common practice of substituting an LLM
judge for the intended user. Had we relied on the
external judge instead of the authors, it would have crowned the most expensive arm, and our headline
ranking of the three arms would have reversed. Gemini ran on a fast web tier (Section~\ref{sec:deviations}), so being the external judge is
confounded here with being a lighter model. We offer the reversal to illustrate what such a substitution can do. It does not pin down how a frontier external judge would order the arms.

The main way these tools scale is by using more computation. Across the three arms the tokens spent on a fixed-length report rose by a factor of about thirty;
the usefulness authors assigned to the output did not. The arms bundle model family, internal
prompting, and the number of calls (Section~\ref{sec:tools}), so this is the return to spending more on the whole configuration rather than to compute in isolation. The scarce input is the author's
hour, and compute is cheap. At the common length the reports imposed roughly similar reading loads, and the extra computation behind the costlier reports did not raise what that hour
returned. The
reversal under the external judge is a warning about measuring that return. When a model stands
in for the intended user, the measured return to that spending can change sign. A journal or a
department weighing such tools for its authors should therefore ask for evidence from the people
who would have to act on the output.

\emph{paper-workshop}'s second act, which we did not score here, returns a tracked-changes revision and a re-run of the
analysis rather than a report, and a redline is a different object from a referee's letter. A study
that asked authors to accept or reject specific edits would measure something we did not. When the
question is which report an author finds most useful, the plain single pass outranked the tools
built to beat it.

\section{Conclusion}
\label{sec:conclusion}

Across a sample of 55 economics meta-analyses, authors returned rankings for 44 papers. In those
rankings, the reports from the single-pass configuration ranked detectably ahead of those from two multi-agent tools that cost far more to run, and the tools were ours. We had pre-registered the opposite expectation, that debate
would help. The lesson we take is about the returns to these more elaborate configurations. More agents, more rounds, and more tokens are costs that have to earn their place, and in this setting we detected no gain in the authors' rankings.

For a researcher deciding how to get a second read on a draft of this kind, the sensible default is to start with the single-pass configuration (this design does not identify the occasions on which the more
elaborate tools would pay). We have not shown that multi-agent debate fails. Cost capped \emph{paper-workshop} at a light setting, and normalization imposed a common length and template, so part of the result may reflect those constraints rather than debate itself. In our own use, outside this
controlled comparison, we still reach for both tools and find them useful (that is our subjective experience, not evidence, and nothing in the rankings
obliges the reader to share it). Three extensions would test the
boundary directly. First, debate may help more when the base model is weaker. Much of what debate
once added was plausibly a second model catching the first's hallucinations and unsupported claims. A strong
model can now do more of that inside a single pass, reasoning over and checking its own draft, so the room for
an external cross-check may narrow as base models improve. Whether a single model checks itself well enough to
account for our null is unsettled, but a weaker-model comparison would test it. Second, we held every
report to a common length. A multi-agent protocol that is concise by design, rather than shortened
afterward to fit, might fare differently. Third, the workshop could run at full depth, which we could
not afford 55 times.

AI tools meant to improve research should be tested the way
other meta-scientific reforms are tested: registered before the data arrive and run against the simplest credible alternative, then scored by their intended users. The code, the blinded reports on our own papers, the rankings, the analysis, the
judge prompts, and the pre-registration are archived so that the measurement
can be repeated or extended. The two tools evaluated here carry citable identifiers of their own
\citep{tool_mad,tool_workshop}. In this design, the Gemini judge, as we ran it, was not a reliable substitute for the intended user.
The one AI judge external to all three reports agreed with the authors least, and relying on it to
rank the three arms would have reversed the finding.

\bigskip
\noindent\textbf{Acknowledgments.} We are grateful to the authors who read the reports and returned a ranking:
Robbie van Aert, Amar Anwar, Anton Astakhov, Laure de Batz, Graziella Bonanno, Petr Cala,
Michael Chletsos, Sefa Awaworyi Churchill, Quirin Dammerer, Dominika Ehrenbergerova,
Ali Elminejad, Alexandra Ferreira-Lopes, Mattia Filomena, Sebastian Gechert, Thomas de Graaff,
Zuzana Gric, Mojmir Hampl, Philipp Heimberger, Stefan Hirsch, Roman Horvath,
Tersoo David Iorngurum, Karel Janda, Evzen Kocenda, Katerina Kroupova, Ludwig List,
Martina Luskova, Simona Malovana, Colin Mang, Satoshi Mizobata, Jiri Novak,
Matej Opatrny, Matteo Picchio, Giorgio Di Pietro, Bob Reed, Miriam Rehm, Jelena Reljic,
Matthias Schnetzer, Andreas Sintos, Michele Ubaldi, Petra Valickova, Tomas Vlach,
Gang Xiao, Xindong Xue, Fan Yang, Ayaz Zeynalov, Olesia Zeynalova, and Diana Zigraiova.
We thank them for their time and their candor. Being named here indicates only that an author returned
a ranking; we do not attach any name to a ranking, comment, or quotation. We also thank the respondent whose
concern we quote anonymously. Any remaining errors are our own.

\medskip
\noindent\textbf{Competing interests.} The authors of this paper developed two of the three AI tools evaluated
here (\emph{mad-research} and \emph{paper-workshop}) as well as related open tools for
multi-agent debate and research auditing (\emph{research-audit-duel-protocol} and
\emph{erc-ai-feedback}). The
pre-registered, identity-masked design was chosen in part because of this conflict. The result runs against that interest: the papers' authors ranked both \emph{mad-research} and \emph{paper-workshop} below the plain single pass. Neither author ranked any paper, including their own.

\medskip
\noindent\textbf{Data availability.} The study was pre-registered on the Open Science Framework
(\href{https://doi.org/10.17605/OSF.IO/E6XGW}{OSF registration e6xgw}); the online supplement is also
available on the project's OSF page (\href{https://osf.io/7nfyb}{osf.io/7nfyb}). A replication package is
archived on Zenodo (DOI \href{https://doi.org/10.5281/zenodo.21273528}{10.5281/zenodo.21273528}). It contains the
de-identified rankings, the analysis code, the AI-judge prompts, the normalization and blinding
specification, the recruitment email templates, the pre-analysis plan, the cost inputs and generated
tables, and the blinded AI reports on our own papers (three per paper) with their letter-to-arm keys. The real journal referee reports,
verbatim author replies, contact data, and the AI reports on external authors' papers are not released. The tool
versions are archived with citable identifiers: \emph{research-audit-duel-protocol}
(\href{https://doi.org/10.5281/zenodo.19105954}{10.5281/zenodo.19105954}), \emph{mad-research}
(\href{https://doi.org/10.5281/zenodo.20829175}{10.5281/zenodo.20829175}), \emph{paper-workshop}
(\href{https://doi.org/10.5281/zenodo.20828996}{10.5281/zenodo.20828996}), and \emph{erc-ai-feedback}
(\href{https://doi.org/10.5281/zenodo.20829165}{10.5281/zenodo.20829165}).

\clearpage
\noindent\textbf{Consent and ethics.} The participants were the papers' authors. They were invited, were told that the attached reports were AI-generated, and could return a one-line
ranking or ignore the invitation. The released participant data are de-identified rankings, not reply
emails or comments; because paper IDs remain, we do not describe them as fully anonymous. The AI
reports released in the archive are study stimuli generated from already public papers, not
participant responses. Real journal referee reports, contact data, and verbatim emails are not
released. The archive honors any participant opt-outs from ranking or report release,
and named acknowledgments indicate only that an author returned a ranking.

\medskip
\noindent\textbf{Use of AI.} The AI tools studied here generated the reports that are the object of
this paper. The authors also used AI coding assistants (Claude Code and the OpenAI Codex CLI) to help build the analysis
pipeline and edit the manuscript. Every study-result number derives from the
frozen analysis pipeline, and the cost and token figures derive from the documented cost memo
and run-time telemetry described in the supplement. The authors are responsible for the final text.

\clearpage
\appendix
\renewcommand{\thetable}{A\arabic{table}}
\setcounter{table}{0}
\section{The sample}
\begingroup
\setlength{\parskip}{0pt}
\begin{longtable}{@{}llll@{}}
\caption{The 55 meta-analyses in the sample.}\label{tab:corpus}\\
\toprule
ID & Stratum & Response & Reference \\
\midrule
\endfirsthead
\multicolumn{4}{c}{\tablename\ \thetable\ (continued)}\\
\toprule
ID & Stratum & Response & Reference \\
\midrule
\endhead
\bottomrule
\endlastfoot
own01 & Own      & Ranked   & \citet{c_own01} \\
own02 & Own      & Ranked   & \citet{c_own02} \\
own03 & Own      & Ranked   & \citet{c_own03} \\
own04 & Own      & Ranked   & \citet{c_own04} \\
own05 & Own      & Ranked   & \citet{c_own05} \\
own06 & Own      & Ranked   & \citet{c_own06} \\
own07 & Own      & Ranked   & \citet{c_own07} \\
own08 & Own      & Ranked   & \citet{c_own08} \\
own09 & Own      & Ranked   & \citet{c_own09} \\
own10 & Own      & Ranked   & \citet{c_own10} \\
own11 & Own      & No ranking & \citet{c_own11} \\
own12 & Own      & Ranked   & \citet{c_own12} \\
own13 & Own      & No ranking & \citet{c_own13} \\
own14 & Own      & Ranked   & \citet{c_own14} \\
own15 & Own      & No ranking & \citet{c_own15} \\
own16 & Own      & Ranked   & \citet{c_own16} \\
own17 & Own      & Ranked   & \citet{c_own17} \\
own18 & Own      & Ranked   & \citet{c_own18} \\
own19 & Own      & Ranked   & \citet{c_own19} \\
own20 & Own      & No ranking & \citet{c_own20} \\
own21 & Own      & No ranking & \citet{c_own21} \\
own22 & Own      & Ranked   & \citet{c_own22} \\
own23 & Own      & Ranked   & \citet{c_own23} \\
own24 & Own      & Ranked   & \citet{c_own24} \\
own25 & Own      & Ranked   & \citet{c_own25} \\
own26 & Own      & Ranked   & \citet{c_own26} \\
own27 & Own      & No ranking & \citet{c_own27} \\
ext01 & External & No ranking & \citet{c_ext01} \\
ext02 & External & Ranked   & \citet{c_ext02} \\
ext03 & External & Ranked   & \citet{c_ext03} \\
ext04 & External & Ranked   & \citet{c_ext04} \\
ext05 & External & Ranked   & \citet{c_ext05} \\
ext06 & External & Ranked   & \citet{c_ext06} \\
ext07 & External & Ranked   & \citet{c_ext07} \\
ext08 & External & No ranking & \citet{c_ext08} \\
ext09 & External & Ranked   & \citet{c_ext09} \\
ext10 & External & Ranked   & \citet{c_ext10} \\
ext11 & External & Ranked   & \citet{c_ext11} \\
ext12 & External & Ranked   & \citet{c_ext12} \\
ext13 & External & Ranked   & \citet{c_ext13} \\
ext14 & External & No ranking & \citet{c_ext14} \\
ext15 & External & Ranked   & \citet{c_ext15} \\
ext16 & External & Ranked   & \citet{c_ext16} \\
ext17 & External & Ranked   & \citet{c_ext17} \\
ext18 & External & No ranking & \citet{c_ext18} \\
ext19 & External & Ranked   & \citet{c_ext19} \\
ext20 & External & Ranked   & \citet{c_ext20} \\
ext21 & External & Ranked   & \citet{c_ext21} \\
ext22 & External & Ranked   & \citet{c_ext22} \\
ext23 & External & Ranked   & \citet{c_ext23} \\
ext24 & External & Ranked   & \citet{c_ext24} \\
ext25 & External & Ranked   & \citet{c_ext25} \\
ext26 & External & Ranked   & \citet{c_ext26} \\
ext27 & External & Ranked   & \citet{c_ext27} \\
ext28 & External & No ranking & \citet{c_ext28} \\
\end{longtable}
\endgroup

\vspace{-1ex}
{\footnotesize
\noindent\emph{Notes:} Response = whether at least one author returned a ranking
within the window. ``No ranking'' includes non-response and one explicit decline,
not identified in the table. The 28 external papers all appear in the
\emph{Journal of Economic Surveys}, either in a published issue or online first.\par}

\clearpage
\bibliography{refs}

\end{document}